\documentclass{article}

\usepackage[preprint]{spconf}
\copyrightnotice{979-8-3503-0689-7/23/\$31.00~\copyright2023 IEEE}

\usepackage{cite,bm}
\usepackage{spconf, amsmath, graphicx}
\usepackage[table]{xcolor}
\usepackage{multirow}
\usepackage{amssymb}
\usepackage{amsfonts}
\usepackage{comment}
\usepackage{amsmath,mathtools}
\usepackage{booktabs}
\usepackage[noend]{algpseudocode}
\usepackage{flushend}
\usepackage{siunitx}
\usepackage{xcolor}
\usepackage{arydshln}
\usepackage{url}
\usepackage[hidelinks,bookmarks=true,hypertexnames=true]{hyperref}
\usepackage{caption} 
\renewenvironment{thebibliography}[1]{
  \begin{oldthebibliography}{#1}
  \setlength{\itemsep}{0.em}
  \setlength{\parskip}{0.em}
}
{
  \end{oldthebibliography}
}

\makeatletter
\def\bstctlcite{\@ifnextchar[{\@bstctlcite}{\@bstctlcite[@auxout]}}
\def\@bstctlcite[#1]#2{\@bsphack
  \@for\@citeb:=#2\do{%
    \edef\@citeb{\expandafter\@firstofone\@citeb}%
    \if@filesw\immediate\write\csname #1\endcsname{\string\citation{\@citeb}}\fi}%
  \@esphack}
\makeatother

\def\L{{\mathcal L}}

\title{A Single Speech Enhancement Model Unifying Dereverberation, Denoising, Speaker Counting, Separation, and Extraction}

\name{
    \!\!\!\!\!\!\!\!\!\!Kohei Saijo$^{1,3}$,
    Wangyou Zhang$^{2,3}$,
    Zhong-Qiu Wang$^{3}$,
    Shinji Watanabe$^{3}$,
    Tetsunori Kobayashi$^{1}$,
    Tetsuji Ogawa$^{1}$
}
\address{
    \hspace{-2em}$^1$Waseda University, Japan \; $^2$Shanghai Jiao Tong University, China \; $^3$ Carnegie Mellon University, USA
}

\begin{document}
\bstctlcite{IEEEexample:BSTcontrol} %
\ninept
\maketitle

\begin{abstract}
\noindent We propose a \textit{multi-task universal speech enhancement} (MUSE) model that can perform five speech enhancement (SE) tasks: dereverberation, denoising, speech separation (SS), target speaker extraction (TSE), and speaker counting.
This is achieved by integrating two modules into an SE model: 1) an internal separation module that does both speaker counting and separation; and 2) a TSE module that extracts the target speech from the internal separation outputs using target speaker cues.
The model is trained to perform TSE if the target speaker cue is given and SS otherwise.
By training the model to remove noise and reverberation, we allow the model to tackle the five tasks mentioned above with a single model, which has not been accomplished yet.
Evaluation results demonstrate that the proposed MUSE model can successfully handle multiple tasks with a single model.
\end{abstract}


\vspace{-1em}
\section{Introduction}
\label{sec:intro}
\vspace{-0.5em}
Speech enhancement (SE), including dereverberation, denoising, speech separation (SS), and target speaker extraction (TSE), has been used as front-ends of speech systems such as automatic speech recognition (ASR) to recover speech corrupted by interference speech and background noises~\cite{haeb2020far}.
In recent years, deep neural networks (DNNs) have led to remarkable progress in SE~\cite{luo2019conv, subakan2021attention}.
Even with only a single microphone, SE models achieve great performance under challenging noisy-reverberant conditions~\cite{wang2023tf}.

For front-end speech processing in multi-talker scenarios, two major approaches are SS~\cite{hershey2016deep, yu2017permutation} and TSE~\cite{vzmolikova2019speakerbeam, wang19h_interspeech, zmolikova2023neural}.
SS aims to separate all the speakers in the mixture, while TSE extracts only the target speech by utilizing additional information about the target speaker, such as enrollment utterances.
TSE is more suitable for applications where the identification of the target speaker is required alongside the separation process. 
However, TSE models cannot work without the target speaker's information, while SS models can work without any additional information.
In~\cite{ochiai2019unified}, unifying TSE and SS systems into a single model was proposed to capitalize on the complementary nature of these two approaches. 
The model first \textit{internally} separates each speaker and then chooses the target speaker feature from internal separation results using an attention-based speaker selection module.
By switching between binary attention (for SS) and soft attention (for TSE), both SS and TSE tasks can be unified within a single model.  
Despite the elegant integration achieved through this approach, it however does not inherit one of the key advantages offered by TSE, and the model cannot handle mixtures with a variable number of speakers as the number of outputs in the internal separation module is fixed.

Several attempts have been made to perform speaker counting along with SS to handle mixtures with a variable number of speakers~\cite{hershey2016deep, nachmani2020voice, zhu2021multi, kolbaek2017multitalker, luo2018speaker, luo20b_interspeech, wisdom2021s}.
One major approach, recursive SS~\cite{kinoshita2018listening, takahashi19_interspeech, neumann20_interspeech, shi2020sequence}, separates speakers one by one in an iterative manner until all the speakers are separated.
Recent work~\cite{chetupalli2023speaker} introduces \textit{internal} separation using the encoder-decoder-based attractor generation (EDA) module, which is initially proposed for speaker diarization~\cite{horiguchi2022encoder}, for efficient and high-performing separation.
Since these methods are also capable of handling a single-speaker input, they perform denoising, SS, and speaker counting with a single model~\cite{chetupalli2023speaker}.
However, they still only address a limited range of SE tasks.

In this paper, we present a \textit{multi-task universal speech enhancement} (MUSE) model that performs five SE tasks: dereverberation, denoising, speaker counting, SS, and TSE.
The proposed model is designed to handle mixtures with a variable number of speakers, including single-speaker scenarios, and it outputs all the speech signals while estimating the number of speakers.
Additionally, the model can extract only the target speech when an enrollment utterance of the target speaker is provided.
To achieve this, we integrate two modules into an SE model: 1) an EDA-based internal separation module that performs both speaker counting and separation; and 2) a TSE module that extracts the target speaker feature from the internal separation outputs.
To efficiently and stably train the proposed model, we introduce a two-stage training strategy, where we first train the modules used for SS and speaker counting and then train the TSE module.
By training the model to remove noise and reverberation, we allow the model to tackle the five tasks mentioned above.
Our experiments on simulated mixtures with noises, reverberation, and up to five speakers show that the proposed MUSE model can handle all five tasks with promising results.
The implementations\footnote{\url{https://github.com/kohei0209/espnet/tree/muse/egs2/wsj0_Nmix}} and the audio samples\footnote{\url{https://kohei0209.github.io/muse-demo/}} are publicly available.

\vspace{-1em}
\section{Multi-task Universal Speech Enhancement}
\label{sec: proposed_method}
\vspace{-0.5em}
\begin{figure}[t]
\centering
\centerline{\includegraphics[width=\linewidth]{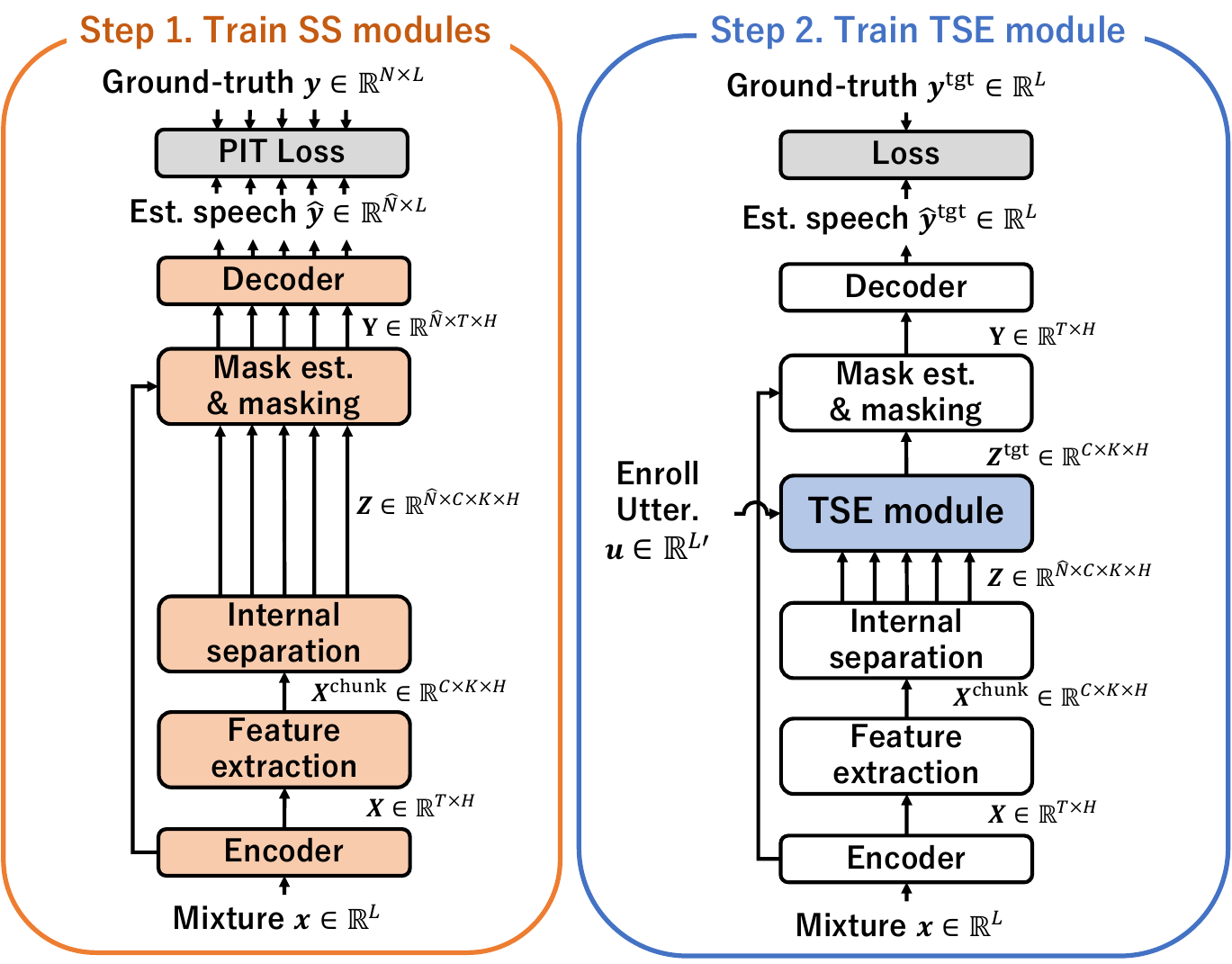}}
\vspace{-2.5mm}
\caption{
    Illustration of proposed model and two-stage training procedure.
    Modules for speaker counting and separation are first trained (left), and then TSE modules are inserted and fine-tuned (right).
    Blocks in white are frozen.
}
\label{fig: overview}
\vspace{-1.5em}
\end{figure}

This paper aims at building a multi-task universal speech enhancement (MUSE) model that performs dereverberation, denoising, speaker counting, SS, and TSE\footnote{Differently from previous studies~\cite{Universal-Kavalerov2019, Universal-Serra2022}, we define \textit{universality} as the capability to handle multiple SE tasks.}.
To achieve this, we exploit two modules: 1)~an \textit{internal separation module} that performs speaker counting and SS (Sect.~\ref{ssec: eda}); and 2)~a \textit{TSE module} that extracts a specified target speaker (Sect.~\ref{ssec: tse}).
By incorporating these two modules into an SE model, the model can handle speaker counting, SS, and TSE.
In addition, by training the model to perform dereverberation and denoising, the model can address all the five tasks mentioned above.

\vspace{-1em}
\subsection{System overview}
\label{ssec: basic_framework}
\vspace{-0.5em}

The proposed model is shown in Fig.~\ref{fig: overview}.
It consists of an \textbf{encoder}, a \textbf{feature extraction module}, an \textbf{internal separation module}, a \textbf{TSE module}, a \textbf{mask estimation module}, and a \textbf{decoder}.
Any SE model architecture can be used for the MUSE model, and we employ the dual-path transformer network (DPTNet)~\cite{chen20l_interspeech} as a basic architecture considering its high performance and compact model size.
As shown in Fig.~\ref{fig: overview}, we first train the modules for speaker counting and separation (shown in the left subplot), and then TSE modules are inserted and fine-tuned (shown in the right subplot).
This subsection provides an overview of each module.

\noindent{\textbf{Encoder:}}
Let us denote a monaural noisy-reverberant mixture containing $N$ speech sources as $\bm{x}\in\mathbb{R}^{L}$, with $L$ denoting the number of samples in the time domain.
The encoder, consisting of a 1-D convolution layer followed by a ReLU activation, maps the time-domain mixture to a feature representation $\bm{X}\in\mathbb{R}^{T \times H}$, where $T$ is the number of frames and $H$ the hidden dimension.

\noindent{\textbf{Feature extraction:}}
The feature extraction module consists of a global layer normalization (gLN)~\cite{luo2019conv}, a segmentation operation, and four DPT blocks.
Following~\cite{luo2020dual}, the encoder output $\bm{X}$ is first normalized with gLN and then split into $C$ chunks along time-frame dimension with a length of $K$ frames in each chunk and an overlap ratio of 50\% between consecutive chunks, resulting in $\bm{X}^{\mathrm{chunk}} \in \mathbb{R}^{C \times K \times H}$ which is then processed by the four DPT blocks.

\noindent{\textbf{Internal separation:}}
The internal separation module consists of an encoder-decoder-based attractor generation (EDA) module~\cite{horiguchi2022encoder} and a DPT block.
The EDA module produces an estimate of the number of speakers (denoted as $\hat{N}$) and a feature representation $\bm{Z}_{n} \in \mathbb{R}^{C \times K \times H}$ for each speaker $n\in\{1,\ldots,\hat{N}\}$, by taking $\bm{X}^{\mathrm{chunk}}$ as the input.
The details of the EDA module will be described later in Sect.~\ref{ssec: eda}.
Next, each $\bm{Z}_{n}$ is processed by a DPT block with its parameters shared for all the $\hat{N}$ estimated speakers.

\noindent{\textbf{TSE module:}}
The TSE module extracts only the target speaker feature $\bm{Z}^{\mathrm{tgt}} \in \mathbb{R}^{C \times K \times H}$ from the $\hat{N}$ internal separation outputs by using an enrollment utterance as a cue.
One of our key contributions is to train our model so that the TSE module is only used when the model is informed to perform TSE and is not used otherwise.
The TSE module will be detailed in Sect.~\ref{ssec: tse}.

\noindent{\textbf{Mask estimation and masking:}}
The mask estimation module has the same architecture as the one after the 6\textsuperscript{th} DPT block of DPTNet.
The output of the internal separation module or the TSE module is first processed by a single DPT block, followed by a parametric ReLU (PReLU) activation and a point-wise 2-D convolution, and then by a chunk-level overlap-add operation to reverse the segmentation operation \cite{chen20l_interspeech}.
The resulting tensor has a shape of ${\hat{N} \times T \times H}$.
Next, the tensor is used to estimate a mask for each target speaker of interest, and the mask is point-wisely multiplied with the encoder output, resulting in a feature representation $\bm{Y}_{n} \in \mathbb{R}^{T \times H}$ for each target speaker of interest $n$.

\noindent{\textbf{Decoder:}}
The decoder is a transposed convolution layer used to reconstruct target speech $\hat{\bm{y}}_{n} \in \mathbb{R}^{L}$ for speaker $n$.

\vspace{-1em}
\subsection{Internal separation module}
\label{ssec: eda}
\vspace{-0.5em}
\begin{figure}[t]
\centering
\centerline{\includegraphics[width=0.7\linewidth]{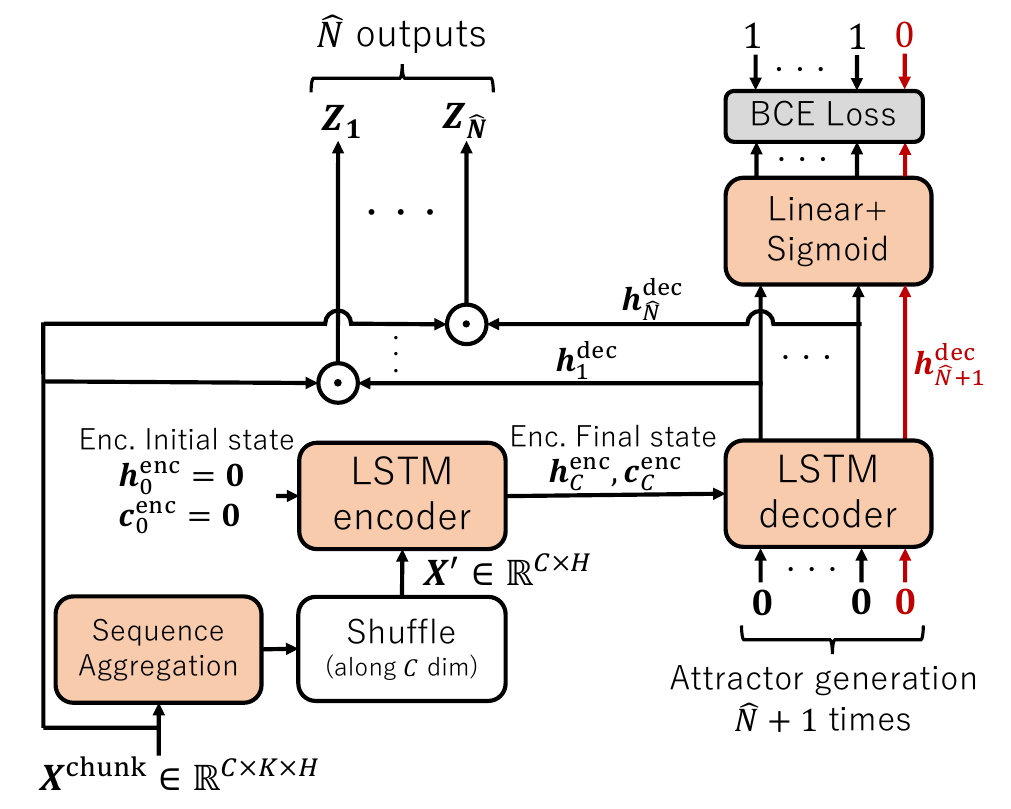}}
\vspace{-3mm}
\caption{
    Overview of EDA in internal separation module.
    EDA estimates an attractor $\bm{h}^{\mathrm{dec}}_{n}$ for each speaker, which is multiplied with input.
    A classifier (Linear + Sigmoid) is trained to decide when to stop attractor generation.
}
\label{fig: eda}
\vspace{-1.5em}
\end{figure}

The internal separation module has an EDA module and a DPT block.
This subsection describes the EDA module, which enables the model to perform speaker counting and SS.
The overview of the EDA module is illustrated in Fig.~\ref{fig: eda}.

\vspace{-1em}
\subsubsection{Attractor generation}
\label{sssec:attractor_generation}
\vspace{-0.5em}
The EDA module contains an intra-chunk sequence aggregation module and an LSTM encoder-decoder, as in~\cite{chetupalli2023speaker}.
We first aggregate the input $\bm{X}^{\mathrm{chunk}} \in \mathbb{R}^{C \times K \times H}$ within each chunk into $\bm{X}' \in \mathbb{R}^{C \times H}$ by the weighted averaging scheme, following~\cite{chetupalli2023speaker}.
$\bm{X}'$ is then shuffled along the chunk dimension to prevent EDA from exploiting the input order information for the attractor generation and speaker counting.
Next, we input $\bm{X}'$, which has a sequence length of $C$, into the LSTM encoder (Fig.~\ref{fig: eda}) and update its hidden state $\bm{h}^{\mathrm{enc}}_{c}$ and cell state $\bm{c}^{\mathrm{enc}}_{c}$ for $C$ times:
\begin{align}
  \label{eqn:lstm_enc}
  \setlength\abovedisplayskip{0pt plus 3pt minus 7pt}
  \setlength\belowdisplayskip{11pt plus 3pt minus 7pt}
    \bm{h}^{\mathrm{enc}}_{c}, \bm{c}^{\mathrm{enc}}_{c} = \mathrm{LSTM}^{\mathrm{enc}}(\bm{X}_{c}', \bm{h}^{\mathrm{enc}}_{c-1}, \bm{c}^{\mathrm{enc}}_{c-1}),
\end{align}
where $c\!\in\!\{1,\!\ldots,\!C\}$ is the chunk index and $\bm{h}^{\mathrm{enc}}_{0} = \bm{c}^{\mathrm{enc}}_{0} = \bm{0}$.
The LSTM decoder (Fig.~\ref{fig: eda}) estimates speaker-wise attractors $\bm{h}^{\mathrm{dec}}_{n}\!\in \!\mathbb{R}^{H}$:
\begin{align}
  \label{eqn:lstm_dec}
    \bm{h}^{\mathrm{dec}}_{n}, \bm{c}^{\mathrm{dec}}_{n} = \mathrm{LSTM}^{\mathrm{dec}}(\bm{0}, \bm{h}^{\mathrm{dec}}_{n-1}, \bm{c}^{\mathrm{dec}}_{n-1}),
\end{align}
where $\bm{h}^{\mathrm{dec}}_{0}$ and  $\bm{c}^{\mathrm{dec}}_{0}$ are respectively initialized using the final states of the LSTM encoder (i.e., $\bm{h}^{\mathrm{dec}}_{0} = \bm{h}^{\mathrm{enc}}_{C}$ and $\bm{c}^{\mathrm{dec}}_{0} = \bm{c}^{\mathrm{enc}}_{C}$).
The estimated attractors are point-wisely multiplied with the input of the EDA module $\bm{X}^{\mathrm{chunk}}$ for each estimated speaker $n$, and in total we obtain $\hat{N}$ speaker representations, denoted as $\bm{Z}_{n} \in \mathbb{R}^{C \times K \times H}$ for speaker $n$.

\vspace{-1em}
\subsubsection{Speaker counting}
\label{sssec:speaker_count}
\vspace{-0.5em}
Our speaker counting module follows \cite{horiguchi2022encoder}.
During training, the EDA module is supervised by the oracle number of speakers $N$ to generate $\hat{N}=N$ attractors.
At run time, we need a criterion to stop the attractor generation for speaker counting.
We use a classifier consisting of a linear layer and a sigmoid activation (shown in the upper right of Fig.~\ref{fig: eda}) to estimate the \textit{attractor existence probability} $p_{n} \in [0, 1]$ for each attractor.
The $n$-th speaker is deemed absent when $p_{n}$ is smaller than a pre-defined threshold and we stop the attractor generation.
The classifier is trained to output ones for the first $N$ iterations and a zero at the $(N\!+\!1)$-th iteration using the binary cross-entropy (BCE) loss, as shown in Fig.~\ref{fig: eda}:
\begin{align}
  \label{eqn:bce_loss}
    \L^{\mathrm{eda}} = \frac{1}{N+1}\mathrm{BCE}(\bm{p}, \hat{\bm{p}}),
\end{align}
where $\bm{p} = [1, \ldots, 1, 0]^\top$ contains $N$ ones and a zero and $\hat{\bm{p}} = [p_1, \ldots, p_{N+1}]^\top$ are the estimated probabilities. 
At run time, we stop the attractor generation process once the estimated probability becomes smaller than 0.5 (i.e., stop at the $({\hat{N}+1})$-th iteration), and use the first $\hat{N}$ attractors to obtain $\bm{Z}_{1},\ldots,\bm{Z}_{\hat{N}}$.

\vspace{-1.1em}
\subsection{TSE module}
\label{ssec: tse}
\vspace{-0.5em}
\begin{figure}[t]
\centering
\centerline{\includegraphics[width=0.65\linewidth]{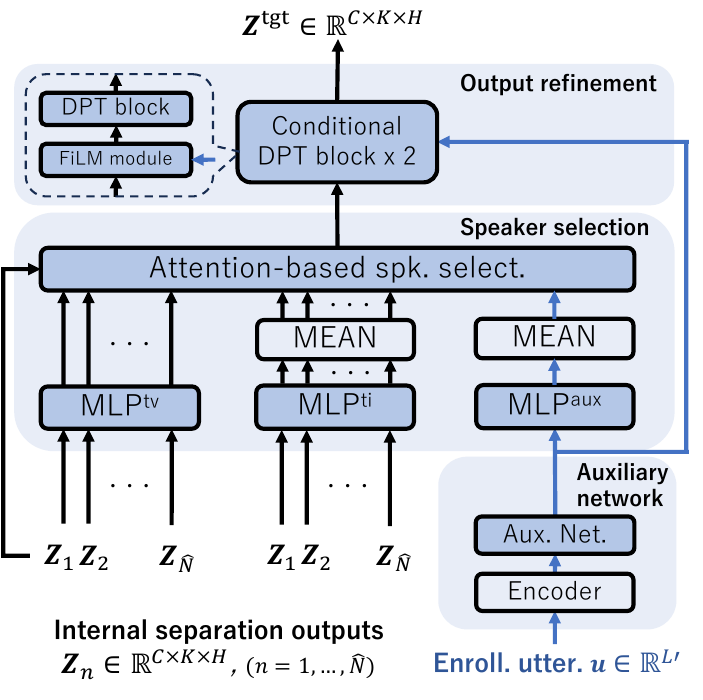}}
\vspace{-4mm}
\caption{
    Overview of TSE module.
    It first computes a target speaker embedding from an enrollment utterance and selects the target speaker feature from internal separation outputs.
    Then selected target speaker feature is further enhanced by two conditional DPT blocks (i.e., output refinement module). 
}
\label{fig: tse}
\vspace{-1.5em}
\end{figure}

This subsection describes the TSE module, which enables the model to perform TSE along with other SE tasks, as introduced in Sect.~\ref{ssec: basic_framework}.
The TSE module, shown in Fig.~\ref{fig: tse}, contains an auxiliary network, a speaker selection module, and an output refinement module.

\vspace{-1.3em}
\subsubsection{Auxiliary network}
\label{sssec:auxiliary_network}
\vspace{-0.5em}
The auxiliary network adopts the same architecture as the feature extraction module in Sect.~\ref{ssec: basic_framework} except that the number of the DPT blocks is two.
Let $\bm{u} \in \mathbb{R}^{L'}$ be the enrollment utterance with $L'$ samples in the time domain.
$\bm{u}$ is first processed using the encoder which is shared with the separation model.
Then we obtain the segmented target speaker embedding $\bm{U}^{\mathrm{chunk}} \in \mathbb{R}^{C' \times K \times H}$ with $C'$ chunks using the auxiliary network in the same way as the feature extraction module.

\vspace{-1.2em}
\subsubsection{Speaker selection module}
\label{sssec:speaker_selection_module}
\vspace{-0.5em}
The speaker selection module aims at selecting the target speaker feature from the $\hat{N}$ internal separation outputs $\{\bm{Z}_{1},\ldots,\bm{Z}_{\hat{N}}\}$ (Sect.~\ref{sssec:attractor_generation}).
We follow the basic idea in~\cite{ochiai2019unified} that introduced an attention-based speaker selection module to select the target speaker feature.
Let us denote the time-variant attention weight of the $n$-th speaker as $\bm{a}_{n} \in [0, 1]^{C \times K \times 1}$.
The speaker selection output is
{\setlength\abovedisplayskip{2pt plus 0pt minus 5pt}
\setlength\belowdisplayskip{2pt plus 0pt minus 5pt}
\begin{align}
  \label{eqn:spk_select}
    \bm{Z}^{\mathrm{tgt}} = \sum\nolimits_{n=1}^{\hat{N}} \bm{a}_{n}\bm{Z}_{n} \in \mathbb{R}^{C \times K \times H}.
\end{align}
}The attention weight is computed using the outputs of internal separation $\bm{Z}_{n}$ and the auxiliary target speaker embedding $\bm{U}^{\mathrm{chunk}}$.
We first obtain time-varying (tv) embeddings $\bm{S}^{\mathrm{tv}}_{n} \in \mathbb{R}^{C \times K \times H'}$ and time-invariant (ti) embeddings $\bm{s}^{\mathrm{ti}}_{n} \in \mathbb{R}^{H'}$ using simple MLP-based networks with $H'$ hidden dimension: $\bm{S}^{\mathrm{tv}}_{n} = \mathrm{MLP}^{\mathrm{tv}}(\bm{Z}_{n}), \; \bm{s}^{\mathrm{ti}}_{n} = \mathrm{MEAN}(\mathrm{MLP}^{\mathrm{ti}}(\bm{Z}_{n}))$, where $\mathrm{MEAN}(\cdot)$ denotes the averaging operation over intra- and inter-chunk dimensions (i.e., $K$ and $C$).
The MLP-based network consists of two linear layers, ReLU activation, and another linear layer.
We also obtain the time-invariant embedding from the auxiliary target speaker embedding $\bm{e}^{\mathrm{aux}} = \mathrm{MEAN}(\mathrm{MLP}^{\mathrm{aux}}(\bm{U}')) \in \mathbb{R}^{H'}$.
Finally, the attention weight is computed as
\begin{align}
  \label{eqn:attention}
    \bm{a}_{n} &= \frac{\mathrm{exp}(\bm{d}_{n})}{\sum\nolimits_{n=1}^{\hat{N}}\mathrm{exp}(\bm{d}_{n})}, \\
    \bm{d}_{n} &= \bm{w}^\top\mathrm{Tanh}(\bm{W}^{\mathrm{tv}}\bm{S}^{\mathrm{tv}}_{n} + \bm{W}^{\mathrm{ti}}\bm{s}^{\mathrm{ti}}_{n} + \bm{W}^{\mathrm{aux}}\bm{e}^{\mathrm{aux}} + \bm{b}),
\end{align}
where $\bm{w}, \bm{W}^{\mathrm{tv}}, \bm{W}^{\mathrm{ti}}, \bm{W}^{\mathrm{aux}}, \bm{b}$ are learnable weights and biases.

\vspace{-1em}
\subsubsection{Output refinement module}
\label{sssec:output_refinement_module}
\vspace{-0.5em}
Since the speaker selection module in Sect.~\ref{sssec:speaker_selection_module} just selects the target speaker features ${\bm{Z}^\mathrm{tgt}}$ using the auxiliary target speaker information, the TSE performance heavily relies on the internal separation performance.
To compensate for the errors from the internal separation, we introduce an output refinement module after the speaker selection module.
This output refinement module consists of two \textit{conditional} DPT blocks composed of a feature-wise linear modulation (FiLM)~\cite{perez2018film} module followed by a DPT block (Fig.~\ref{fig: tse}).
The FiLM module consists of a linear layer, a PReLU activation, a FiLM layer, and another linear layer.
The FiLM layer integrates the auxiliary target speaker embedding into the speaker selection output $\bm{Z}^{\mathrm{tgt}}$:
\begin{align}
  \label{eqn:film}
    \bm{Z}^{\mathrm{tgt}} \leftarrow \gamma(\bm{u}')\bm{Z}^{\mathrm{tgt}} + \beta(\bm{u}'),
\end{align}
where $\gamma(\cdot)$ and $\beta(\cdot)$ are linear layers, and $\bm{u}' = \mathrm{MEAN}(\bm{U}') \in \mathbb{R}^{H}$.
The output of the FiLM module is further processed by a normal DPT block for refinement, as shown in Fig.~\ref{fig: tse}.

\vspace{-1em}
\subsection{Two-stage training}
\label{ssec:training}
\vspace{-0.5em}
We found that training all the modules of the model at the same time causes instability, especially in challenging conditions (e.g., $N \geq 3$).
This is likely because the speaker selection part in the TSE module becomes unstable when the internal separation module does not perform sufficiently well, especially in the first several training epochs.
To deal with this, we propose a two-stage training strategy, where we first train the modules other than the TSE module and then train only the TSE module, as shown in the left and right subplots of Fig.~\ref{fig: overview}.

In the first stage, we train the encoder, feature extraction module, internal separation module, mask estimation module, and decoder.
Given the $N$ estimated speech signals $\hat{\bm{y}} \in \mathbb{R}^{N \times L}$ from the model, we compute the permutation-invariant loss~\cite{yu2017permutation} with reference speech signals ${\bm{y}} \in \mathbb{R}^{N \times L}$ as
\begin{align}
  \label{eqn:pit_loss}
    \L^{\rm{pit}} = \min_{\bm{P}} \frac{1}{N} \sum\nolimits_{n=1}^{N} {\L(\bm{y}_{n}, [\bm{P}\hat{\bm{y}}]_{n})},
\end{align}
where $\L$ is a signal-level loss function and $\bm{P}$ is an ${N \times N}$ permutation matrix.
The gradients are computed using the sum of the loss for separation in Eq.(\ref{eqn:pit_loss}) and speaker counting in Eq.(\ref{eqn:bce_loss}) as
\begin{align}
  \label{eqn:total_loss}
    \L^{\rm{sep}} = \L^{\rm{pit}} + \L^{\rm{eda}}.
\end{align}

In the second stage, we insert the TSE module between the internal separation and mask estimation modules, as shown in Fig.~\ref{fig: overview} (right).
We train only the TSE module while freezing the other parameters.
The loss is computed between the estimated target speech $\hat{\bm{y}}^{\mathrm{tgt}} \in \mathbb{R}^{L}$ and the reference $\bm{y}^{\mathrm{tgt}} \in \mathbb{R}^{L}$ as
\begin{align}
  \label{eqn:tse_loss}
    \L^{\rm{tse}} = {\L(\bm{y}^{\mathrm{tgt}}, \hat{\bm{y}}^{\mathrm{tgt}})}.
\end{align}
Note that, different from stage 1, in stage 2 we only compute $\L^{\rm{tse}}$ and do not compute $\L^{\rm{eda}}$ because the EDA module is not updated.

\subsection{Relation to prior work}
\label{ssec: relations}

\noindent{\textbf{SS with speaker counting based on EDA: }}
In~\cite{chetupalli2023speaker}, EDA-based speaker counting, originally designed for speaker diarization~\cite{horiguchi2022encoder}, is introduced and shown to be capable of performing denoising, separation, and speaker counting with a single model.
Although our MUSE model is based on the same EDA-based method, there are several differences:
1)~we enable the model to handle TSE by introducing a carefully designed TSE module;
2)~our model performs dereverberation while the model in~\cite{chetupalli2023speaker} does not; and 
3)~we use not only fully-overlapped mixtures, which are often used to evaluate models that perform speaker counting~\cite{nachmani2020voice, shi2020sequence, chetupalli2023speaker}, but also partially-overlapped mixtures for evaluation. This allowed us to evaluate WER and conduct deeper analysis on SS with speaker counting.

\noindent{\textbf{Integration of SS and TSE: }}
In~\cite{ochiai2019unified}, a speaker selection module is introduced inside the separation model to unify SS and TSE.
Although the speaker selection module in our TSE module is motivated by the method in \cite{ochiai2019unified}, we introduce several techniques to improve versatility and performance:
1)~our MUSE model can handle a variable number of speakers by leveraging the EDA module, while the model in~\cite{ochiai2019unified} works only for mixtures with a fixed number of speakers;
2)~we introduce a two-stage training strategy, which leads to more stable training, while SS and TSE modules are trained at the same time in~\cite{ochiai2019unified};
and 3)~we introduce an output refinement module to the TSE module (Fig.~\ref{fig: tse}), which can compensate for the drawback of the speaker selection module caused by its heavy reliance on the internal separation performance.

\vspace{-1em}
\section{Experiments}
\label{sec:experiments}
\vspace{-0.5em}

\begin{table*}[t]
\begin{center}
\caption{
    Results on \textbf{anechoic} WSJ0-mix test set. 
    Input, Baseline, and MUSE columns are performances of unprocessed mixtures, models trained with only $N$-mix data, and proposed MUSE model trained using 2-, 3-, 4-, and 5-mix data combined, respectively.
    In MUSE$^\star$, proposed MUSE model estimated the number of speakers. Note that same model checkpoint was used for evaluation in MUSE and MUSE$^\star$.
}
\label{table:results_anechoic}
\resizebox{\linewidth}{!}{
\begin{tabular}{cc|rrrr|rrrr|rrrr|rrr}
\toprule
& & \multicolumn{4}{c|}{SI-SNR [dB]} & \multicolumn{4}{c|}{SDR [dB]} & \multicolumn{4}{c|}{NB-PESQ} & \multicolumn{3}{c}{WER [\%]} \\
\midrule
Task& Test set & Input & Baseline & MUSE & MUSE$^\star$ & Input & Baseline & MUSE & MUSE$^\star$ & Input & Baseline & MUSE & MUSE$^\star$ & Input & MUSE & MUSE$^\star$ \\
\midrule
&2-mix &0.0 &\textbf{18.8} &18.2 &18.2 &0.2 &\textbf{19.2} &18.7 &18.7 &1.68 &\textbf{3.65} &3.58 &3.58 &62.5  &\textbf{7.5} &\textbf{7.5}  \\
&3-mix &-3.2 &12.8 &\textbf{13.9} &\textbf{13.9} &-2.9 &13.9 &\textbf{14.5} &\textbf{14.5} &1.42 &2.87 &\textbf{2.99} &\textbf{2.99} &98.1   &\textbf{10.1} &\textbf{10.1}  \\
&4-mix &-5.0 &8.3 &\textbf{9.5} &\textbf{9.5} &-4.7 &9.1 &\textbf{10.2} &\textbf{10.2} &1.34 &2.26 &\textbf{2.44} &\textbf{2.44} &132.4   &\textbf{22.2}  &22.9  \\
\multirow{-4}{*}{SS}
& 5-mix &-6.3 &5.5 &\textbf{6.5}  &6.4  &-5.8 &6.4 &\textbf{7.4}  &7.3  &1.30 &1.91 &\textbf{2.09} &\textbf{2.09} &175.1  &\textbf{35.0} &35.9 \\

\midrule

& 2-mix &0.0  &15.4 &\textbf{17.6} &\textbf{17.6} &0.2  &16.4 &\textbf{18.1} &\textbf{18.1} &1.68 &3.43 &\textbf{3.55} &\textbf{3.55} &62.5 &\textbf{17.6} &\textbf{17.6} \\
& 3-mix &-3.2 &11.0 &\textbf{12.7} &\textbf{12.7} &-2.9 &11.9 &\textbf{13.5} &\textbf{13.5} &1.42 &2.77 &\textbf{2.94} &\textbf{2.94} &98.1 &\textbf{25.9} &\textbf{25.9} \\
& 4-mix &-5.0 &6.4 &\textbf{7.4} &\textbf{7.4} &-4.7 &7.3 &\textbf{8.4} &\textbf{8.4} &1.34 &2.19 &\textbf{2.35} &\textbf{2.35} &132.4 &\textbf{48.1} &\textbf{48.1} \\
\multirow{-4}{*}{TSE}
& 5-mix &-6.3 &4.0 &\textbf{4.5} &\textbf{4.5} &-5.8 &5.0 &\textbf{5.6} &5.5 &1.30 &1.87 &\textbf{2.03} &\textbf{2.03} &175.1  &67.2 &\textbf{67.0} \\

\bottomrule
\end{tabular}
}
\end{center}
\vspace{-2em}
\end{table*}

\begin{table*}[t]
\begin{center}
\caption{
    Results on \textbf{noisy-reverberant} WSJ0-mix test set.
    Input, baseline, and MUSE columns respectively present the results of unprocessed mixtures, models trained with only $N$-mix data, and the proposed MUSE model trained using 1-, 2-, 3-, 4-, and 5-mix data combined.
    In MUSE$^\star$, proposed MUSE model estimated the number of speakers.
    Note that same model checkpoint was used for evaluation in MUSE and MUSE$^\star$.
}
\label{table:results_noisyrev}
\resizebox{\linewidth}{!}{
\begin{tabular}{cc|rrrr|rrrr|rrrr|rrr}
\toprule
& & \multicolumn{4}{c|}{SI-SNR [dB]} & \multicolumn{4}{c|}{SDR [dB]} & \multicolumn{4}{c|}{NB-PESQ} & \multicolumn{3}{c}{WER [\%]} \\
\midrule
Task& Test set & Input & Baseline & MUSE & MUSE$^\star$ & Input & Baseline & MUSE & MUSE$^\star$ & Input & Baseline & MUSE & MUSE$^\star$ & Input & MUSE & MUSE$^\star$ \\

\midrule

& 1-mix &4.5   &\textbf{11.5} &\textbf{11.5} &\textbf{11.5} &6.7   &13.8  &\textbf{13.9} &\textbf{13.9} &2.09 &3.00 &\textbf{3.01} &\textbf{3.01} &\textbf{9.2} &10.4 &10.4  \\
& 2-mix &-1.6  &8.2  &\textbf{8.4}  &\textbf{8.4}  &-0.8  &9.9  &\textbf{10.1} &\textbf{10.1} &1.59  &2.43 &\textbf{2.45} &\textbf{2.45} &69.1 &\textbf{21.9} &22.3  \\
& 3-mix &-4.6  &4.9  &\textbf{5.4}  &5.3  &-3.7  &6.5  &\textbf{7.0}  &6.9  &1.42  &1.99 &\textbf{2.05} &\textbf{2.05} &104.8 &\textbf{45.5} &47.2  \\
& 4-mix &-6.4  &2.2  &\textbf{2.5}  &2.4  &-5.4  &3.9  &\textbf{4.1}  &4.0  &1.35  &1.76 &\textbf{1.80} &1.79 &122.0 &\textbf{71.1} &73.8  \\
\multirow{-5}{*}{SS}
& 5-mix &-7.7  &0.0  &\textbf{0.3}  &0.2  &-6.7  &1.8  &\textbf{1.9}  &\textbf{1.9}  &1.32  &1.62 &\textbf{1.64} &\textbf{1.64} &148.5 &\textbf{94.2} &94.6  \\

\midrule

& 1-mix &4.5  &11.5 &\textbf{11.6} &\textbf{11.6} &6.7 &13.8 &\textbf{13.9} &\textbf{13.9} &2.09 &3.00 &\textbf{3.02} &\textbf{3.02} &\textbf{9.2} &10.3 &10.3  \\
& 2-mix &-1.6 &6.9 &\textbf{7.8}   &\textbf{7.8}  &-0.8 &8.6 &\textbf{9.6}  &9.5 &1.59 &2.35 &\textbf{2.42} &\textbf{2.42} &69.1 &\textbf{31.2} &31.7  \\
& 3-mix &-4.6 &3.6 &\textbf{4.4}   &\textbf{4.4}  &-3.7 &5.3 &\textbf{6.0}  &\textbf{6.0}  &1.42 &1.96 &\textbf{2.01} &\textbf{2.01} &104.8 &\textbf{58.6} &59.5  \\
& 4-mix &-6.4 &0.4 &\textbf{0.9}   &0.8  &-5.4 &2.0 &\textbf{2.6}  &2.5  &1.35 &1.70 &\textbf{1.74} &\textbf{1.74} &122.0 &89.0 &\textbf{88.0}  \\
\multirow{-5}{*}{TSE}
& 5-mix &-7.7 &-1.6  &\textbf{-1.3} &-1.4 &-6.7 &0.1 &\textbf{0.4} &\textbf{0.4} &1.32 &1.57 &\textbf{1.60} &\textbf{1.60}  &148.5 &\textbf{111.1} &111.3  \\

\bottomrule

\end{tabular}
}
\end{center}
\vspace{-2.5em}
\end{table*}

\vspace{-1em}
\subsection{Dataset}
\vspace{-0.5em}
To evaluate the effectiveness of the proposed method for multiple tasks, we utilized the WSJ0-mix dataset\footnote{\url{https://github.com/mpariente/pywsj0-mix}}.
It contained a \textit{min} version of $2$-, $3$-, $4$-, and $5$-speaker mixtures simulated using clean speech in the WSJ0 corpus~\cite{wsj0}.
Each $N$-speaker mixture ($N=2,3,4,5$) contained $20000$, $5000$, and $3000$ mixtures for training, validation, and testing, respectively.
The sampling rate of all data was $8$\si{\kilo\hertz}.

To extend the proposed method for a noisy-reverberant setup, we synthesized a noisy-reverberant version of the WSJ0-mix dataset by including noises and room reverberation\footnote{We used a customized dataset because there is no open-source noisy-reverberant dataset for 3-, 4-, and 5-speaker mixtures. The simulation script is included in the published training recipe.}.
The noises were drawn from the WHAM! dataset~\cite{wham} and the noise gain is adjusted to attain an SNR between \SIrange{0}{15}{\decibel} against the speaker with the lowest power in each mixture.
Room impulse responses (RIRs) were simulated with Pyroomacoustics~\cite{pra} and the reverberation times ranged between \SIrange{150}{650}{\milli\second}.
We convolved the early part (first \SI{50}{\milli\second}) of RIRs with the clean speech as in~\cite{kinoshita2013reverb} to obtain the reference signal. 
In the noisy-reverberant setup, we also simulated 1-speaker mixtures
to train the model to handle single-speaker cases.
The 1-speaker dataset contained 8769, 3553, and 1770 mixtures for training, validation, and testing, respectively.

\vspace{-1em}
\subsection{Training details}
\vspace{-0.5em}
We chose DPTNet~\cite{chen20l_interspeech} as the backbone of our model.
In the encoder/decoder, we set the kernel size and shift to \SI{2}{\milli\second} and \SI{1}{\milli\second}, respectively.
The number of filters $H$ was set to 64.
The model had six DPT blocks in total:
four in the feature extraction module, one in the internal separation module, and one in the mask estimation module.
Note that the TSE module had additional two conditional DPT blocks with $H=64$ as described in Sect.~\ref{sssec:output_refinement_module}.
The hidden dimension of the speaker selection module $H'$ was set to 512 (Sect.\ref{sssec:speaker_selection_module}).

All the experiments were done using the ESPNet toolkit~\cite{espnet, ESPnet_SE-Li2021}.
During training, the batch size was 4 and the input was 4 seconds long.
Scale-invariant signal-to-noise ratio (SI-SNR)~\cite{sisdr} was used as the loss function $\L$ in Eqs.(\ref{eqn:pit_loss}) and (\ref{eqn:tse_loss}).
The model was trained using the Adam optimizer~\cite{adam}.
The learning rate was increased linearly from $0$ to $L_r$ over the first $20000$ training steps and then decayed by $0.98$ every two epochs.
In the anechoic condition, we trained models for $175$ epochs and the peak learning rate $L_r$ was $4\times 10^{-4}$.
In the noisy-reverberant setup, we initialized the model using the one trained with the anechoic data and trained for $100$ epochs with $L_r=10^{-4}$.
In both conditions, the TSE module was trained from scratch for 50 epochs with $10000$ warm-up steps and $L_r=10^{-4}$

\vspace{-1em}
\subsection{Evaluation details}
\vspace{-0.5em}
For evaluation, we considered four metrics, SI-SNR, signal-to-distortion ratio (SDR)~\cite{Performance-Vincent2006, fast_bss_eval}, narrow-band perceptual evaluation of speech quality (NB-PESQ)~\cite{pesq}, and word error rate (WER)\footnote{To evaluate WER, we used the \textit{max} version data. The SS and TSE modules were fine-tuned with \textit{max} data for 10 and 5 epochs, respectively. The separated signals are upsampled to 16\si{\kilo\hertz} before fed into the ASR model.}.
The Whisper Large v2 model\footnote{\url{https://huggingface.co/openai/whisper-large-v2}}~\cite{Whisper-Radford2022} was used for evaluating the ASR performance with greedy decoding.

We evaluated the performance with and without assuming the knowledge of the oracle number of speakers.
In the latter case, the number of estimated speakers $\hat{N}$ can be different from the oracle number $N$ since the model conducted speaker counting using the classifier (Sect.~\ref{sssec:speaker_count}).
When the model \textit{overestimated} the number (i.e., $\hat{N} > N$), we computed the optimal assignment of estimate-reference pairs and ignored the rest $\hat{N} - N$ speakers.
In contrast, when the model \textit{underestimated} the number (i.e., $\hat{N} < N$), we first computed the optimal assignment of estimate-reference pairs, and then compensated for underestimated speeches by copying one of the $\hat{N}$ estimated signals that gave the highest SDR with the reference speech for each of the remaining $N - \hat{N}$ references.

\vspace{-1em}
\subsection{Results on anechoic setup}
\vspace{-0.5em}

\begin{figure}[t]
\centering
\centerline{\includegraphics[width=0.9\linewidth]{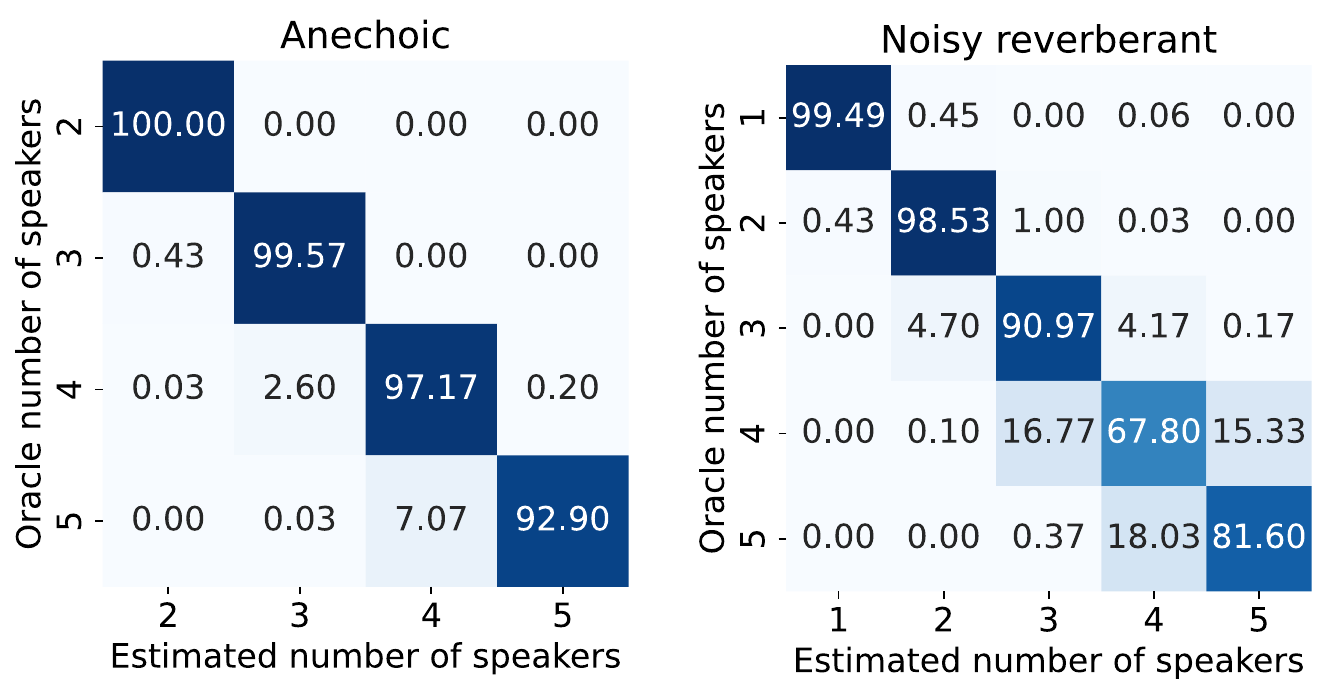}}
\vspace{-3mm}
\caption{
    Confusion matrix of speaker counting rate [\%] in anechoic (left) and noisy reverberant (right) conditions. Numbers in diagonal cells correspond to speaker counting accuracy.
}
\label{fig:confusion_matrix}
\vspace{-1.5em}
\end{figure}

We first examine the effectiveness of the proposed MUSE framework using the \textbf{anechoic} data.
Here, the model handles three tasks: SS, TSE, and speaker counting.
In addition to the proposed model, we trained several baseline models with the same architecture using only the 2-, 3-, 4-, or 5-mix data to verify whether the performance of the proposed method is reasonable\footnote{We trained each baseline for the same number of epochs as the MUSE model. Although this is not a completely fair comparison, the purpose of comparing with the baseline model is to see whether the proposed method achieves reasonable performance, not to show which is superior.}, and report the results in the columns denoted as ``Baseline''.
From Table~\ref{table:results_anechoic}, we observe that the MUSE model successfully handled both SS and TSE on all 2-5 mix test sets.
Significant improvement in both speech separation and recognition metrics indicates the effectiveness of the model as a front-end for diverse applications such as hearing aids and ASR.
Furthermore, it is worth noting that the model is capable of counting the number of speakers with more than 90\% accuracy for all 2-5 mix data, as shown in Fig.~\ref{fig:confusion_matrix} (left). 
In particular, the model estimated the number of speakers almost perfectly on 2- or 3-mix data.
It is also observed that both SE and ASR performances degrade when the number of speakers in the mixture increases, which is in line with the results reported in~\cite{chetupalli2023speaker}.
Overall, we confirm that the proposed MUSE model can successfully handle SS, TSE, and speaker counting with high performance.

Interestingly, we found that the performance is similar whether the model was given the oracle speaker number $N$ or not (MUSE versus MUSE$^\star$).
This means that the qualities of the $N - \hat{N}$ separated signals obtained by enforcing the model to estimate $N$ sources are low, considering the fact that most counting error comes from under-estimation.
We attribute this to the nature of the EDA.
Let us focus on the case where the model underestimates the speaker number. %
As described in Sect.~\ref{ssec: eda}, the classifier in the EDA is trained to estimate a low existence probability for the $(N+1)$-th attractor, which is not used for separation.
In other words, separation performance when using an attractor with a low probability is not necessarily great.
However, such attractors with low existence probabilities are used when the oracle number of speakers is given.
Thus, the qualities of the last $N - \hat{N}$ separated signals can be poor.

\vspace{-1em}
\subsection{Results on noisy-reverberant setup}
\vspace{-0.5em}
After verifying the performance on anechoic conditions, we extend the MUSE model to additionally handle dereverberation and denoising by exploiting the \textbf{noisy-reverberant} data.
Table~\ref{table:results_noisyrev} shows the results.
First of all, even in the challenging noisy-reverberant setup, the model can still handle both SS and TSE tasks.
Additionally, the result on 1-mix data demonstrates that the proposed model can successfully address the denoising and dereverberation tasks in the single-speaker scenario.
On the other hand, the performance improvement was limited compared to the anechoic case, especially on the challenging 4- or 5-mix data.
The results in Table~\ref{table:results_noisyrev} and Fig.~\ref{fig:confusion_matrix} (right) suggest that the reverberation and noise make both separation and speaker counting more difficult.
While we observed some performance degradation compared to anechoic conditions, our study is the first in the literature to report SE and ASR performance under realistic noisy reverberant conditions, encompassing up to five SE tasks.
We hope this result and our released codes can attract more attention to tackle multiple SE tasks for variable numbers of speakers in realistic scenarios.
In future work, we plan to leverage a larger amount of data to tackle challenging conditions and take advantage of the proposed MUSE model to leverage more diverse data regardless of the tasks or the number of speakers.

\vspace{-1em}
\section{Conclusions}
\label{sec:conclusions}
\vspace{-0.5em}
This paper has made the first attempt towards building a universal SE model that can handle all of the dereverberation, denoising, speaker counting, SS, and TSE tasks.
The EDA module enables the model to perform speaker counting along with SS, and the proposed TSE module addresses TSE without any performance degradation in speaker counting and SS.
Evaluation results demonstrate that the proposed MUSE model can address speaker counting, SS, and TSE with high performance in anechoic conditions and has strong potential in challenging noisy-reverberant conditions.

\vspace{-0.5em}
\section{ACKNOWLEDGMENTS}
\label{sec:ack}
\vspace{-0.5em}
This work was supported in part by JST SPRING, Grant Number JPMJSP2128.
Part of the experiments were done using the PSC Bridges2 system via ACCESS allocation CIS210014, supported by National Science Foundation grants \#2138259, \#2138286, \#2138307, \#2137603, and \#2138296.

\newpage
{%
\bibliographystyle{IEEEtran}
\bibliography{main}}

\end{document}